# MASSLESS $XXZ$ MODEL AND DEGENERATION OF THE ELLIPTIC ALGEBRA $\mathcal{A}_{Q,P}\left(\widehat{SL}_2\right)$

MICHIO JIMBO

*Department of Mathematics, Faculty of Science, Kyoto University, Kyoto 606, Japan.*

HITOSHI KONNO

*The Division of Mathematical and Information Sciences, Faculty of Integrated Arts and Sciences, Hiroshima University, Higashi-Hiroshima 739, Japan.*

AND

TETSUJI MIWA

*Research Institute for Mathematical Sciences, Kyoto University, Kyoto 606, Japan.*

**Abstract.** We consider an algebraic structure of the $XXZ$ model in the gapless regime. We argue that a certain degeneration limit of the elliptic algebra $\mathcal{A}_{q,p}(\widehat{sl}_2)$ [5] is a relevant object. We give a free boson realization of this limiting algebra and derive an integral formula for the correlation function. The result agrees with the one obtained by solving a system of difference equations. We also discuss the relation of our algebra to the deformed Virasoro algebra and Lukyanov's bosonization of the sine-Gordon theory.

## 1. Introduction

The one-dimensional spin 1/2 XXZ chain is a system defined by the Hamiltonian

$$H = -\frac{1}{2} \sum_{n=-\infty}^{\infty} \left( \sigma_n^x \sigma_{n+1}^x + \sigma_n^y \sigma_{n+1}^y + \Delta \sigma_n^z \sigma_{n+1}^z \right). \qquad (1.1)$$

Here the $\sigma_i^\alpha$ ($\alpha = x, y, z$) stand for the usual Pauli matrices acting on the $i$-th lattice site. We are concerned with the case $|\Delta| < 1$, where the spectrum of (1.1) is gapless. In a recent work [1], the problem of describing correlation functions of this model was discussed. Let us first quote the result below.

By a correlation function we mean the ground-state average $\langle \mathcal{O} \rangle$ of a local operator
$$\mathcal{O} = E^{(i_1)}_{\varepsilon_1 \varepsilon'_1} \cdots E^{(i_k)}_{\varepsilon_k \varepsilon'_k}.$$

Here $E_{\varepsilon \varepsilon'} = (\delta_{a\varepsilon} \delta_{b\varepsilon'})_{a,b=\pm}$ is the matrix unit, and $E^{(i)}_{\varepsilon \varepsilon'}$ denotes the operator acting as $E_{\varepsilon \varepsilon'}$ on the $i$-th site and as identity elsewhere. It suffices to consider $\mathcal{O} = E^{(1)}_{\varepsilon'_1 \varepsilon_1} \cdots E^{(n)}_{\varepsilon'_n \varepsilon_n}$, since the general case is reduced to this case by translation and taking linear combinations. Set

$$\Delta = -\cos \pi \nu \qquad (0 < \nu < 1). \tag{1.2}$$

Define the indices $\bar{1}, \cdots, \bar{n}$ ($1 \leq \bar{1} < \cdots < \bar{n} \leq 2n$) from $\varepsilon_i, \varepsilon'_i$ by the rule

$$\{\bar{1}, \cdots, \bar{n}\} = \{j \mid 1 \leq j \leq n, \varepsilon'_j = -\} \cup \{j^* \mid 1 \leq j \leq n, \varepsilon_j = +\}$$

where $j^* = 2n + 1 - j$. With these notations, the formula written down in [1] reads as follows:

$$\begin{aligned}
&\langle E^{(1)}_{\varepsilon'_1 \varepsilon_1} \cdots E^{(n)}_{\varepsilon'_n \varepsilon_n} \rangle \\
&= \frac{1}{\nu^{n(n-1)/2}} \int_{-\infty}^{\infty} \cdots \int_{-\infty}^{\infty} \prod_{l=1}^{n} \frac{d\alpha_l}{2\pi} \prod_{1 \leq l < l' \leq n} \frac{\sinh(\alpha_l - \alpha_{l'})}{\sinh \nu(\alpha_l - \alpha_{l'} - \pi i)} \\
&\quad \times \prod_{1 \leq \bar{l} \leq n} \Big[ \prod_{j=1}^{n} \frac{i}{\sinh(\alpha_l + i0)} \prod_{1 \leq j < \bar{l}} \sinh \nu(\alpha_l) \prod_{\bar{l} < j \leq n} \sinh \nu(-\alpha_l + \pi i) \Big] \\
&\quad \times \prod_{n+1 \leq \bar{l} \leq 2n} \Big[ \prod_{j=1}^{n} \frac{-i}{\sinh(\alpha_l - i0)} \prod_{1 \leq j < \bar{l}^*} \sinh \nu(-\alpha_l) \prod_{\bar{l}^* < j \leq n} \sinh \nu(\alpha_l + \pi i) \Big].
\end{aligned} \tag{1.3}$$

This is a rather unwieldy, yet completely explicit formula for correlation functions of an arbitrary local operator.

In the earlier works [2, 3], the case of the anti-ferromagnetic regime $\Delta < -1$ was treated. Being written solely in terms of elementary functions, the formula (1.3) is simpler than the corresponding one for the massive case $\Delta < -1$. Nevertheless the situation in the massless case $|\Delta| < 1$ has remained unclear for quite some time. The main reason is that the corner transfer matrix argument is not directly applicable in the massless regime. In the work [1], (1.3) was derived through the following indirect argument.

The massless XXZ model is a limiting case of the XYZ model. The latter is massive, and the corner transfer matrix method applies. This leads to the following system of difference equations.

$$G_n(\cdots,\beta_{j+1},\beta_j,\cdots)_{\cdots,\varepsilon_{j+1},\varepsilon_j,\cdots}$$
$$= \sum_{\varepsilon'_j,\varepsilon'_{j+1}} R^{\varepsilon'_j,\varepsilon'_{j+1}}_{\varepsilon_j,\varepsilon_{j+1}}(\beta_j - \beta_{j+1})G_n(\cdots,\beta_j,\beta_{j+1},\cdots)_{\cdots,\varepsilon'_j,\varepsilon'_{j+1},\cdots}, \tag{1.4}$$

$$G_n(\beta_1,\cdots,\beta_{2n-1},\beta_{2n} - i\lambda)_{\varepsilon_1,\cdots,\varepsilon_{2n}} = G_n(\beta_{2n},\beta_1,\cdots,\beta_{2n-1})_{\varepsilon_{2n},\varepsilon_1,\cdots,\varepsilon_{2n-1}}, \tag{1.5}$$

$$G_n(\beta_1,\cdots,\beta_{2n})_{\varepsilon_1,\cdots,\varepsilon_{2n}}\bigg|_{\beta_{2n}=\beta_{2n-1}+\pi i}$$
$$= \delta_{\varepsilon_{2n-1}+\varepsilon_{2n},0}G_{n-1}(\beta_1,\cdots,\beta_{2n-2})_{\varepsilon_1,\cdots,\varepsilon_{2n-2}}. \tag{1.6}$$

Here $G_n$ is a function with values in $(\mathbf{C}^2)^{\otimes 2n}$, and $R$ denotes the elliptic $R$ matrix (see (2.3) below). The correlation functions are obtained by taking $\lambda = 2\pi$ and specializing the variables as

$$(\beta_1,\cdots,\beta_{2n}) = (\overbrace{\beta + \pi i,\cdots,\beta + \pi i}^{n},\overbrace{\beta,\cdots,\beta}^{n}).$$

We note that these equations imply in particular the quantum Knizhnik-Zamolodchikov ($q$KZ) equation. In the elliptic case, however, construction of solutions of the above system is still an open problem. The formula (1.3) was obtained by solving the corresponding difference equations directly in the XXZ limit. The difficulty in this approach is that the solutions of difference equations are determined only up to arbitrary periodic functions, so one has to single out in some way the unique solution which corresponds to the correlation functions.[1] For this reason, most part of [1] is devoted to the verification that (1.3) reproduces the correct results in all known cases: the XXX limit ($\nu = 0$), Ising limit ($\nu = 1/2$), and the nearest neighbor two point function ($\nu$ general). Thus it is fair to call (1.3) a conjecture, supported by these evidences.

For the massive XXZ model $\Delta < -1$, the ambiguity of solutions is resolved since the canonical solution is provided by a trace of vertex operators. The latter has a clear meaning in the framework of representation theory of the quantum affine algebra $U_q(\widehat{sl}_2)$, and via bosonization, enables us to write down an explicit integral formula for the trace similar to (1.3). The situation in the massless case would become much clearer if a similar bosonization is found for the full elliptic algebra $\mathcal{A}_{q,p}\left(\widehat{sl}_2\right)$ [5] corresponding to the XYZ model. In this connection, we should mention that Lukyanov [6] had previously constructed a bosonization of the vertex operators for the sine-Gordon model. Although his formulas look quite similar to (1.3), the precise relation has not been worked out. This note is intended to make explicit the connections between the elliptic algebra, Lukyanov's bosonization, and the correlation functions (1.3).

---

[1] Compare with the form factor bootstrap approach [4], in which the abundance of solutions corresponds to that of local operators. In our case there is only one physically meaningful solution.

Lukyanov's construction starts with bosonized vertex operators with a cutoff, and the form factors of the sine-Gordon model are derived after removing the cutoff parameter at the final stage. Here we prefer to work directly with operators with the cutoff parameter removed. The regularization is introduced in the integrals appearing in the contractions of these operators. The vertex operators are written in terms of bosonic oscillators alone, without involving 'zero mode' operators $P, Q$ (see the remark at the end of section 3). We verify that the formula (1.3) for the correlations coincides with the trace of these vertex operators (of type I in the terminology of [3]). The vertex operators considered here can be regarded as a degeneration of those of the elliptic algebra $\mathcal{A}_{q,p}\left(\widehat{sl}_2\right)$ in the limit $p, q \to 1$. The precise relation is given in the text. In particular we show that the symmetry property of the $L$ operators which play a key role in the definition of $\mathcal{A}_{q,p}\left(\widehat{sl}_2\right)$ is indeed satisfied in this limit.

## 2. Degeneration of the elliptic algebra

In this section, we discuss the degeneration of the elliptic algebra $\mathcal{A}_{q,p}\left(\widehat{sl}_2\right)$ and the vertex operators associated with it.

### 2.1. $R$ MATRIX

In order to fix the notation, let us begin by recalling the elliptic $R$ matrix from Baxter's book [7]. The Boltzmann weights of the eight vertex model are given by (see eq.(10.4.21) in [7])

$$a(u) = \frac{\operatorname{snh}(\lambda - u)}{\operatorname{snh}(\lambda)}, \quad b(u) = \frac{\operatorname{snh}(u)}{\operatorname{snh}(\lambda)}, \quad c(u) = 1, \quad d(u) = k\operatorname{snh}(\lambda - u)\operatorname{snh}(u). \tag{2.1}$$

Here $\operatorname{snh}(u) = -i\operatorname{sn}(iu)$, and $\operatorname{sn}(u)$ denotes Jacobi's elliptic function with modulus $k$. Let $K, K'$ be the corresponding complete elliptic integrals. We shall use the variables

$$p = e^{-\frac{\pi K'}{K}}, \qquad q = -e^{-\frac{\pi \lambda}{2K}}, \qquad \zeta = e^{\frac{\pi u}{2K}}, \tag{2.2}$$

and regard (2.1) as functions of $\zeta, p, q$.

The $R$ matrix is defined to be

$$R(\zeta) = R(\zeta; p^{1/2}, q^{1/2}) = \frac{1}{\mu(\zeta)} \begin{pmatrix} a(u) & & & d(u) \\ & b(u) & c(u) & \\ & c(u) & b(u) & \\ d(u) & & & a(u) \end{pmatrix}. \quad (2.3)$$

We choose the overall scalar factor $\mu(\zeta)$ as follows. [2]

$$\frac{1}{\mu(\zeta)} = \frac{1}{\overline{\kappa}(\zeta^2)} \frac{(p^2; p^2)_\infty}{(p;p)_\infty^2} \frac{\Theta_{p^2}(q^2)\Theta_{p^2}(p\zeta^2)}{\Theta_{p^2}(q^2\zeta^2)}, \quad (2.4)$$

$$\frac{1}{\overline{\kappa}(z)} = \frac{(q^4 z^{-1}; p, q^4)_\infty (q^2 z; p, q^4)_\infty (pz^{-1}; p, q^4)_\infty (pq^2 z; p, q^4)_\infty}{(q^4 z; p, q^4)_\infty (q^2 z^{-1}; p, q^4)_\infty (pz; p, q^4)_\infty (pq^2 z^{-1}; p, q^4)_\infty} \quad (2.5)$$

where

$$(z; p_1, \cdots, p_m)_\infty = \prod_{n_1, \cdots, n_m \geq 0} (1 - z p_1^{n_1} \cdots p_m^{n_m}), \quad (2.6)$$

$$\Theta_q(z) = (z; q)_\infty (qz^{-1}; q)_\infty (q; q)_\infty. \quad (2.7)$$

With this choice of $\mu(\zeta)$, (2.3) satisfies the unitarity and crossing relations as well as the Yang-Baxter equation. We skip the details since we do not need them.

As usual, we shall regard (2.3) as a linear map acting on $V \otimes V$ with $V = \mathbf{C}^2$. In the standard basis $v_+, v_-$ of $V$ we write

$$R(\zeta) v_{\varepsilon_1} \otimes v_{\varepsilon_2} = \sum_{\varepsilon_1', \varepsilon_2' = \pm} v_{\varepsilon_1'} \otimes v_{\varepsilon_2'} \, R_{\varepsilon_1 \varepsilon_2}^{\varepsilon_1' \varepsilon_2'}(\zeta).$$

2.2. ELLIPTIC ALGEBRA $\mathcal{A}_{Q,P}\left(\widehat{SL}_2\right)$

The algebra $\mathcal{A}_{q,p}\left(\widehat{sl}_2\right)$ [5, 9] was proposed as an elliptic extension of the quantized affine algebra $U_q\left(\widehat{sl}_2\right)$. It is presented in terms of the formal generating series

$$L^\pm(\zeta) = \sum_{n=-\infty}^{\infty} L_n^\pm \zeta^{-n}, \qquad L_n^\pm = \left(L_{\varepsilon\varepsilon',n}^\pm\right)_{\varepsilon,\varepsilon' = \pm}, \quad (2.8)$$

$$L_{\varepsilon\varepsilon',n}^\pm = 0 \quad \text{if } \varepsilon\varepsilon' \neq (-1)^n. \quad (2.9)$$

---

[2] In this occasion let us correct some misprints in [8]. On page 434, the second line from the bottom, the formula for $d(\zeta)$ should read $k\,\mathrm{snh}(\lambda - u)\mathrm{snh}(u)/\mu(\zeta)$. In the first line of eq.(2.18), the factor $(p^2; p^2)_\infty^2$ should read $(p^2; p^2)_\infty$.

By definition, $\mathcal{A}_{q,p}\left(\widehat{sl}_2\right)$ is the algebra generated by the symbols $L^{\pm}_{\varepsilon\varepsilon',n}$ ($n \in \mathbf{Z}, \varepsilon, \varepsilon' = \pm, \varepsilon\varepsilon' = (-1)^n$) and a central element $c$, through the following defining relations.

$$R^{\pm}_{12}(\zeta_1/\zeta_2) \overset{1}{L^{\pm}}(\zeta_1) \overset{2}{L^{\pm}}(\zeta_2) = \overset{2}{L^{\pm}}(\zeta_2) \overset{1}{L^{\pm}}(\zeta_1) R^{*\pm}_{12}(\zeta_1/\zeta_2), \qquad (2.10)$$

$$R^{+}_{12}(q^{c/2}\zeta_1/\zeta_2) \overset{1}{L^{+}}(\zeta_1) \overset{2}{L^{-}}(\zeta_2) = \overset{2}{L^{-}}(\zeta_2) \overset{1}{L^{+}}(\zeta_1) R^{*+}_{12}(q^{-c/2}\zeta_1/\zeta_2), \qquad (2.11)$$

$$q\text{-det}L^{+}(\zeta) \equiv L^{+}_{++}(q^{-1}\zeta)L^{+}_{--}(\zeta) - L^{+}_{-+}(q^{-1}\zeta)L^{+}_{+-}(\zeta) = q^{c/2}, \qquad (2.12)$$

$$L^{-}_{\varepsilon\varepsilon'}(\zeta) = \varepsilon\varepsilon' L^{+}_{-\varepsilon,-\varepsilon'}(p^{1/2}q^{-c/2}\zeta). \qquad (2.13)$$

Let us explain the notation. We set

$$R^{+}(\zeta) = \tau(q^{1/2}\zeta^{-1})R(\zeta), \qquad R^{-}(\zeta) = \tau(q^{1/2}\zeta)^{-1}R(\zeta),$$

where

$$\tau(\zeta) = \zeta^{-1}\frac{(q\zeta^2;q^4)_\infty(q^3\zeta^{-2};q^4)_\infty}{(q^3\zeta^2;q^4)_\infty(q\zeta^{-2};q^4)_\infty}. \qquad (2.14)$$

We use also the $R$ matrices in which the parameter $p$ is changed:

$$R^{*\pm}(\zeta) = R^{\pm}(\zeta; p^{*1/2}, q^{1/2}), \qquad p^* = pq^{-2c}.$$

Our generating series (2.8) contain both positive and negative powers of $\zeta$. If we set

$$L^{+}_{\varepsilon\varepsilon',n} = \left(-p^{1/2}\right)^{\max(n,0)} \overline{L}^{+}_{\varepsilon\varepsilon',n}$$

and let formally $p \to 0$, then $L^{+}(\zeta)$ and $L^{-}(\zeta)$ become power series in $\zeta$ and $\zeta^{-1}$ respectively. In this limit, the relations (2.10)-(2.13) reduce to the presentation of the quantum affine algebra $U_q(\widehat{sl}_2)$ due to Reshetikhin and Semenov-Tian-Shanskii [10]. As it is argued in [5], the symmetry condition (2.13) means that for generic $p$ we have the same number of generators as we have for $p = 0$.

### 2.3. VERTEX OPERATORS

In the application of the representation theory of $U_q(\widehat{sl}_2)$, an important role is played by the integrable highest weight modules and intertwiners between them (vertex operators). From now on, we shall focus attention to the level one case. In this case, there exist two highest weight modules $\mathcal{H}^{(i)}$ ($i = 0, 1$). The vertex operators are intertwiners of the form

$$\Phi^{(1-i,i)}(\zeta) : \mathcal{H}^{(i)} \longrightarrow \mathcal{H}^{(1-i)} \otimes V_\zeta, \qquad \Phi^{(1-i,i)}(\zeta) = \sum \Phi^{(1-i,i)}_\varepsilon(\zeta) \otimes v_\varepsilon,$$

$$\Psi^{*(1-i,i)}(\zeta) : V_\zeta \otimes \mathcal{H}^{(i)} \longrightarrow \mathcal{H}^{(1-i)}, \qquad \Psi^{*(1-i,i)}_\varepsilon(\zeta) = \Psi^{*(1-i,i)}(\zeta)(v_\varepsilon \otimes \cdot).$$

Here $V_\zeta$ denotes the evaluation module based on $V = \mathbf{C}^2 = \mathbf{C}v_+ \oplus \mathbf{C}v_-$ (see e.g.[3] for the details).

It was conjectured in [5] that in the elliptic case $\mathcal{A}_{q,p}\left(\widehat{sl}_2\right)$ there exist natural analogs of the level one modules and vertex operators. (Here we say that a representation of $\mathcal{A}_{q,p}\left(\widehat{sl}_2\right)$ has level $k$ if the central element $c$ acts as $k$ times the identity.) Let us use the same letters $\mathcal{H}^{(i)}, \Phi_\varepsilon^{(1-i,i)}(\zeta), \Psi_\varepsilon^{*(1-i,i)}(\zeta)$ for the elliptic counterpart. We shall often omit writing the upper indices $i$ when there is no fear of confusion. The conjecture says in particular that they obey the following commutation relations.

$$\Phi_{\varepsilon_2}(\zeta_2)\Phi_{\varepsilon_1}(\zeta_1) = \sum_{\varepsilon_1',\varepsilon_2'=\pm} R_{\varepsilon_1\varepsilon_2}^{\varepsilon_1'\varepsilon_2'}(\zeta_1/\zeta_2)\Phi_{\varepsilon_1'}(\zeta_1)\Phi_{\varepsilon_2'}(\zeta_2), \qquad (2.15)$$

$$\Phi_{\varepsilon_1}(\zeta_1)\Psi_{\varepsilon_2}^*(\zeta_2) = \tau(\zeta_1/\zeta_2)\Psi_{\varepsilon_2}^*(\zeta_2)\Phi_{\varepsilon_1}(\zeta_1), \qquad (2.16)$$

$$\Psi_{\varepsilon_1}^*(\zeta_1)\Psi_{\varepsilon_2}^*(\zeta_2) = -\sum_{\varepsilon_1',\varepsilon_2'=\pm} R_{\varepsilon_1\varepsilon_2}^{*\varepsilon_1'\varepsilon_2'}(\zeta_1/\zeta_2)\Psi_{\varepsilon_2'}^*(\zeta_2)\Psi_{\varepsilon_1'}^*(\zeta_1). \qquad (2.17)$$

In addition, they should satisfy the noramilzation conditions

$$\Phi_{\varepsilon_1}^{(i,1-i)}(\zeta_1)\Phi_{\varepsilon_2}^{(1-i,i)}(\zeta_2) = (-1)^{1-i}\varepsilon_2 g^{-1}\delta_{\varepsilon_1+\varepsilon_2,0} + O(\zeta_1 - q\zeta_2) \quad (\zeta_1 \to q\zeta_2), \qquad (2.18)$$

$$\Psi_{\varepsilon_1}^{*(i,1-i)}(\zeta_1)\Psi_{\varepsilon_2}^{*(1-i,i)}(\zeta_2) = \frac{(-1)^{1-i}\varepsilon_1}{1-q^{-2}\zeta_2^2/\zeta_1^2}\left(\frac{\zeta_2}{q\zeta_1}\right)^{i+(1+\varepsilon_1)/2} g\delta_{\varepsilon_1+\varepsilon_2,0} + O(1)$$
$$(\zeta_1 \to q^{-1}\zeta_2), \qquad (2.19)$$

with some constant $g$.

In terms of the vertex operators, the $L^\pm$ operators acting on $\mathcal{H}^{(i)}$ can be expressed as

$$L_{\varepsilon\varepsilon'}^+(\zeta) = \kappa\Psi_{\varepsilon'}^*(\zeta)\Phi_\varepsilon(q^{1/2}\zeta), \qquad (2.20)$$

$$L_{\varepsilon\varepsilon'}^-(\zeta) = \kappa\Phi_\varepsilon(\zeta)\Psi_{\varepsilon'}^*(q^{1/2}\zeta). \qquad (2.21)$$

Here $\kappa$ is a normalization constant. The defining relations (2.10),(2.11) are immediate consequences of the commutation relations (2.15) -(2.17). The condition (2.12) for the quantum determinant also follows from (2.18),(2.19) with an appropriate choice of $\kappa$. The symmetry (2.13) of the $L$ operators entails the following relation for the vertex operators.

$$\Phi_\varepsilon(\zeta)\Psi_{\varepsilon'}^*(q^{1/2}\zeta) = \varepsilon\varepsilon'\Psi_{-\varepsilon'}^*(p^{1/2}q^{-1/2}\zeta)\Phi_{-\varepsilon}(p^{1/2}\zeta). \qquad (2.22)$$

2.4. DEGENERATION

As we mentioned already, the elliptic algebra degenerates in the limit $p \to 0$ to the quantum affine algebra $U_q(\widehat{sl}_2)$. We now study the degeneration in another limit $p, q \to 1$. More precisely, we let

$$p = q^{2(\xi+1)}, \quad \zeta = q^{i\beta/\pi}, \qquad q \to 1, \tag{2.23}$$

where $\xi$ and $\beta$ are kept fixed.

*Remark.* In [11, 2, 3], the anti-ferroelectric regime of the six vertex model is discussed. This is a limiting case $p \to 0$ of the eight vertex model in the principal regime $0 < p^{1/2} < -q < \zeta^{-1} < 1$. On the other hand, the limit (2.23) is *not* a boundary of the principal regime.

In the limit (2.23), the elliptic $R$ matrix (2.3) degenerates to a trigonometric one. However the limit has the form

$$R = \begin{pmatrix} a & & & d \\ & b & c & \\ & c & b & \\ d & & & a \end{pmatrix}$$

with $d \neq 0$, as opposed to the standard trigonometric $R$ matrix coming from the universal $R$ matrix of $U_q(\widehat{sl}_2)$. In order to bring it to the usual form, we need to introduce a 'gauge' transformation. Define

$$U = \begin{pmatrix} 1 & -i \\ 1 & i \end{pmatrix}, \qquad U_0 = U\,\sigma^z, \quad U_1 = \sigma^z\, U\, \sigma^z.$$

Then we find

$$\lim \left(U_1 \otimes U_0\right) R(\zeta; p^{1/2}, q^{1/2}) \left(U_0 \otimes U_1\right)^{-1} = R(-\beta), \tag{2.24}$$

$$\lim \left(U_1 \otimes U_0\right) \left(-R(\zeta; p^{*1/2}, q^{1/2})\right) \left(U_0 \otimes U_1\right)^{-1} = S(\beta), \tag{2.25}$$

where

$$S(\beta) = \frac{S_0(\beta)}{\sinh\frac{i\pi-\beta}{\xi}} \begin{pmatrix} \sinh\frac{i\pi-\beta}{\xi} & & & \\ & \sinh\frac{\beta}{\xi} & \sinh\frac{i\pi}{\xi} & \\ & \sinh\frac{i\pi}{\xi} & \sinh\frac{\beta}{\xi} & \\ & & & \sinh\frac{i\pi-\beta}{\xi} \end{pmatrix} \tag{2.26}$$

$$R(\beta) = -S(-\beta)|_{\xi \to \xi+1}. \tag{2.27}$$

Here $S_0(\beta)$ is given in Appendix, (A.11). The matrices $\left(S^{cd}_{ab}(\beta)\right)$, $\left(R^{cd}_{ab}(\beta)\right)$ coincide with those given respectively in eq.(2.33), (9.15) of [6].

Note that
$$(\sigma^z \otimes \sigma^z)(U_1 \otimes U_0) = U_0 \otimes U_1,$$
$$(U_0 \otimes U_1)^{-1}(\sigma^z \otimes \sigma^z) = (U_1 \otimes U_0)^{-1}.$$

Since $R(-\beta)$ and $S(\beta)$ commute with $\sigma^z \otimes \sigma^z$, (2.24),(2.25) can also be written as
$$\lim (U_0 \otimes U_1) R(\zeta; p^{1/2}, q^{1/2}) (U_1 \otimes U_0)^{-1} = R(-\beta), \qquad (2.28)$$
$$\lim (U_0 \otimes U_1) \left(-R(\zeta; p^{*1/2}, q^{1/2})\right) (U_1 \otimes U_0)^{-1} = S(\beta). \quad (2.29)$$

The transformations (2.24), (2.25), (2.28), (2.29) are not similarity transformations. This may seem strange at first glance. However, this is understood to be natural because in (2.15) and (2.17) the order of $\zeta_1$ and $\zeta_2$ is opposite in the left and right hand sides.

Now we will rewrite (2.15) and (2.17) in the limit. Write $\Phi_a(\beta), \Psi_a^*(\beta)$ for the limit of $\Phi_a(\zeta), \Psi_a^*(\zeta)$. In accordance with (2.24),(2.25), (2.28), (2.29) we set

$$Z_a^{(1,0)}(\beta) = 2 \sum_b (U_0^{-1})_{ba} \Psi_b^{*(1,0)}(\beta) = \Psi_+^{*(1,0)}(\beta) - ia\Psi_-^{*(1,0)}(\beta) \quad (2.30)$$

$$Z_a^{(0,1)}(\beta) = 2 \sum_b (U_1^{-1})_{ba} \Psi_b^{*(0,1)}(\beta) = a\Psi_+^{*(0,1)}(\beta) - i\Psi_-^{*(0,1)}(\beta) \quad (2.31)$$

$$Z_a'^{(1,0)}(\beta) = \sum_b (U_0)_{ab} \Phi_b^{(1,0)}(\beta) = \Phi_+^{(1,0)}(\beta) + ia\Phi_-^{(1,0)}(\beta), \qquad (2.32)$$

$$Z_a'^{(0,1)}(\beta) = \sum_b (U_1)_{ab} \Phi_b^{(0,1)}(\beta) = a\Phi_+^{(0,1)}(\beta) + i\Phi_-^{(0,1)}(\beta). \qquad (2.33)$$

Their commutation relations can be determined using (2.37),(2.24) and (2.25). Dropping the upper indices, we find

$$Z_a(\beta_1) Z_b(\beta_2) = \sum_{c,d} S_{ab}^{cd}(\beta_1 - \beta_2) Z_d(\beta_2) Z_c(\beta_1), \qquad (2.34)$$

$$Z_a'(\beta_1) Z_b'(\beta_2) = \sum_{c,d} R_{ab}^{cd}(\beta_1 - \beta_2) Z_d'(\beta_2) Z_c'(\beta_1), \qquad (2.35)$$

$$Z_a(\beta_1) Z_b'(\beta_2) = ab \tan\left(\frac{\pi}{4} + i\frac{\beta_1 - \beta_2}{2}\right) Z_b'(\beta_2) Z_a(\beta_1). \qquad (2.36)$$

Here we have used
$$\lim \tau(q^{i\beta/\pi}) = \tan\left(\frac{\pi}{4} + \frac{i\beta}{2}\right). \qquad (2.37)$$

The conditions (2.18),(2.19) become

$$Z_a(\beta_1)Z_b(\beta_2) = \frac{C}{\beta_1 - \beta_2 - \pi i}\delta_{a+b,0} + O(1) \qquad (\beta_1 \to \beta_2 + \pi i), \tag{2.38}$$

$$Z'_a(\beta)Z'_b(\beta + \pi i) = C'\delta_{a+b,0}. \tag{2.39}$$

Here $C, C'$ are constants depending on the normalization of $Z_a(\beta), Z'_a(\beta)$. In addition, the symmetry relation (2.22) reduces to the following.

$$Z'_a(\beta)Z_b(\beta - \frac{\pi i}{2}) = Z_b(\beta^*)Z'_a(\beta^* - \frac{\pi i}{2}), \qquad \beta^* = \beta - \pi i(\xi + \frac{1}{2}). \tag{2.40}$$

In the next section we shall discuss the bosonization of these relations (2.34)-(2.40).

## 3. Bosonization of the vertex operators

In this section we give a bosonization of the operators $Z_\pm(\alpha)$ and $Z'_\pm(\alpha)$.

Let us consider free bosons $a(t)$ $(t \in \mathbf{R})$ which satisfy

$$[a(t), a(t')] = \frac{\sinh\frac{\pi t}{2}\sinh\pi t\sinh\frac{\pi t(\xi+1)}{2}}{t\sinh\frac{\pi t\xi}{2}}\delta(t+t'). \tag{3.1}$$

We also use $a'(t)$ defined by

$$a'(t)\sinh\frac{\pi t(\xi+1)}{2} = a(t)\sinh\frac{\pi t\xi}{2}.$$

We consider the Fock space $\mathcal{H}$ generated by $|\text{vac}\rangle$ which satisfies

$$a(t)|\text{vac}\rangle = 0 \quad \text{if} \quad t > 0.$$

We set

$$V(\alpha) =: e^{i\phi(\alpha)} :, \qquad i\phi(\alpha) = \int_{-\infty}^{\infty}\frac{a(t)}{\sinh\pi t}e^{i\alpha t}dt,$$

$$\overline{V}(\alpha) =: e^{-i\overline{\phi}(\alpha)} :, \qquad i\overline{\phi}(\alpha) = \int_{-\infty}^{\infty}\frac{a(t)}{\sinh\frac{\pi t}{2}}e^{i\alpha t}dt,$$

$$V'(\alpha) =: e^{i\phi'(\alpha)} :, \qquad i\phi'(\alpha) = -\int_{-\infty}^{\infty}\frac{a'(t)}{\sinh\pi t}e^{i\alpha t}dt,$$

$$\overline{V}'(\alpha) =: e^{-i\overline{\phi}'(\alpha)} :, \qquad i\overline{\phi}'(\alpha) = -\int_{-\infty}^{\infty}\frac{a'(t)}{\sinh\frac{\pi t}{2}}e^{i\alpha t}dt.$$

In Appendix 1, we list the operator products of these operators.

A comment is in order here. When we compute the contractions of operators, we often encounter an integral

$$\int_0^\infty F(t)dt$$

which is divergent at $t = 0$. Here we adopt the following prescription for the regularization: it should be understood as the contour integral

$$\int_C F(t)\frac{\log(-t)}{2\pi i}dt, \tag{3.2}$$

where $C$ is the contour

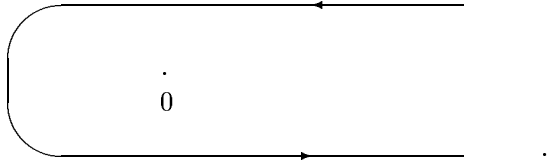

Now let us define

$$Z_+(\beta) = V(\beta), \tag{3.3}$$

$$Z_-(\beta) = \int_{C_1} \frac{d\alpha}{2\pi} : e^{i\phi(\beta)-i\bar{\phi}(\alpha)} : f(\alpha - \beta), \tag{3.4}$$

$$Z'_+(\beta) = V'(\beta), \tag{3.5}$$

$$Z'_-(\beta) = \int_{C_2} \frac{d\alpha}{2\pi} : e^{i\phi'(\beta)-i\bar{\phi}'(\alpha)} : f'(\alpha - \beta), \tag{3.6}$$

where

$$f(\alpha) = \Gamma\left(\frac{i\alpha}{\pi\xi} - \frac{1}{2\xi}\right)\Gamma\left(-\frac{i\alpha}{\pi\xi} - \frac{1}{2\xi}\right),$$

$$f'(\alpha) = \Gamma\left(\frac{i\alpha}{\pi(\xi+1)} + \frac{1}{2(\xi+1)}\right)\Gamma\left(-\frac{i\alpha}{\pi(\xi+1)} + \frac{1}{2(\xi+1)}\right).$$

Here the integration contours are chosen as follows. The contour $C_1$ is $(-\infty, \infty)$ except that the poles

$$\alpha - \beta = -\frac{\pi i}{2} + n\pi\xi i \quad (n \in \mathbf{Z}_{\geq 0})$$

of $\Gamma(\frac{i(\alpha-\beta)}{\pi\xi} - \frac{1}{2\xi})$ are above $C_1$ and the poles

$$\alpha - \beta = \frac{\pi i}{2} - n\pi\xi i \quad (n \in \mathbf{Z}_{\geq 0})$$

of $\Gamma(-\frac{i(\alpha-\beta)}{\pi\xi} - \frac{1}{2\xi})$ are below $C_1$. The contour $C_2$ is $(-\infty, \infty)$. The poles

$$\alpha - \beta = \frac{\pi i}{2} + n\pi(\xi+1)i \quad (n \in Z_{\geq 0})$$

of $\Gamma(\frac{i(\alpha-\beta)}{\pi(\xi+1)} + \frac{1}{2(\xi+1)})$ are above $C_2$ and the poles

$$\alpha - \beta = -\frac{\pi i}{2} - n\pi(\xi+1)i \quad (n \in Z_{\geq 0})$$

of $\Gamma(-\frac{i(\alpha-\beta)}{\pi(\xi+1)} + \frac{1}{2(\xi+1)})$ are below $C_2$. We are going to show that the operators (3.3)-(3.5) satisfy the relations for the degeneration of the elliptic algebra.

**Proposition 3.1** *The operators $Z_\pm(\beta)$ and $Z'_\pm(\beta)$ satisfy the commutation relations (2.34), (2.35) and (2.36).*

*Proof.* First we consider (2.34). The equality

$$Z_+(\beta_1)Z_+(\beta_2) = S_0(\beta_1 - \beta_2)Z_+(\beta_2)Z_+(\beta_1)$$

follows immediately from (A.1).

Note that

$$w(\alpha) = c_1 f(\alpha) \sinh\frac{\alpha - \frac{\pi i}{2}}{\xi}, \quad c_1 = \frac{e^{-(\gamma+\log(\pi\xi))\frac{\xi+1}{\xi}}}{\pi i}.$$

We have

$$\begin{aligned}
Z_+(\beta_1)Z_-(\beta_2) &= c_1 g(\beta_2 - \beta_1)\int_{C_1}\frac{d\alpha}{2\pi} : e^{i\phi(\beta_1)+i\phi(\beta_2)-i\bar{\phi}(\alpha)} : \\
&\times f(\alpha - \beta_1)f(\alpha - \beta_2)\sinh\frac{\alpha - \beta_1 - \frac{\pi i}{2}}{\xi}. \quad (3.7)
\end{aligned}$$

The contour $C_1$ must be properly chosen with respect to the location of the poles of $f(\alpha - \beta_1)$ and $f(\alpha - \beta_2)$. As for $f(\alpha - \beta_2)$ this is already discussed. Since the factor $f(\alpha - \beta_1)\sinh\frac{\alpha-\beta_1-\frac{\pi i}{2}}{\xi}$ is from the contraction of the kind (A.2), the choice of $C_1$ is such that $\text{Im}(\alpha - \beta_1) < -\frac{\pi}{2}$. Note that the zeros of $\sinh\frac{\alpha-\beta_1-\frac{\pi i}{2}}{\xi}$ cancel the poles of $\Gamma(-\frac{i\alpha}{\pi\xi} - \frac{1}{2\xi})$. Therefore, we can keep the same condition on $C_1$ with respect to the poles of $f(\alpha - \beta_1)$ as $f(\alpha - \beta_2)$. Similarly, we have

$$Z_+(\beta_2)Z_-(\beta_1) = c_1 g(\beta_1 - \beta_2)\int_{C_1}\frac{d\alpha}{2\pi} : e^{i\phi(\beta_1)+i\phi(\beta_2)-i\bar{\phi}(\alpha)} :$$

$$\times f(\alpha - \beta_1)f(\alpha - \beta_2)\sinh\frac{\alpha - \beta_2 - \frac{\pi i}{2}}{\xi}, \qquad (3.8)$$

$$Z_-(\beta_2)Z_+(\beta_1) = -c_1 g(\beta_1 - \beta_2)\int_{C_1}\frac{d\alpha}{2\pi} : e^{i\phi(\beta_1)+i\phi(\beta_2)-i\bar{\phi}(\alpha)} :$$
$$\times f(\alpha - \beta_1)f(\alpha - \beta_2)\sinh\frac{\alpha - \beta_1 + \frac{\pi i}{2}}{\xi}, \qquad (3.9)$$

We used (A.12) in the latter equality. The condition on the contour $C_1$ is the same for all (3.7), (3.8), (3.9). Therefore, the equality

$$Z_+(\beta_1)Z_-(\beta_2) = S_0(\beta_1 - \beta_2)\{\frac{\sinh\frac{\pi i}{\xi}}{\sinh\frac{\pi i - \beta_1 + \beta_2}{\xi}}Z_+(\beta_2)Z_-(\beta_1)$$
$$+ \frac{\sinh\frac{\beta_1 - \beta_2}{\xi}}{\sinh\frac{\pi i - \beta_1 + \beta_2}{\xi}}Z_-(\beta_2)Z_+(\beta_1)\}$$

follows from the identity

$$\sinh\frac{\alpha - \beta_1 - \frac{\pi i}{2}}{\xi}\sinh\frac{\pi i - \beta_1 + \beta_2}{\xi}$$
$$= \sinh\frac{\alpha - \beta_2 - \frac{\pi i}{2}}{\xi}\sinh\frac{\pi i}{\xi} - \sinh\frac{\alpha - \beta_1 + \frac{\pi i}{2}}{\xi}\sinh\frac{\beta_1 - \beta_2}{\xi}.$$

The proof of the equality

$$Z_-(\beta_1)Z_+(\beta_2) = S_0(\beta_1 - \beta_2)\{\frac{\sinh\frac{\pi i}{\xi}}{\sinh\frac{\pi i - \beta_1 + \beta_2}{\xi}}Z_-(\beta_2)Z_+(\beta_1)$$
$$+ \frac{\sinh\frac{\beta_1 - \beta_2}{\xi}}{\sinh\frac{\pi i - \beta_1 + \beta_2}{\xi}}Z_+(\beta_2)Z_-(\beta_1)\}$$

is similar. Now, we proceed to the equality

$$Z_-(\beta_1)Z_-(\beta_2) = S_0(\beta_1 - \beta_2)Z_-(\beta_2)Z_-(\beta_1). \qquad (3.10)$$

We have

$$Z_-(\beta_1)Z_-(\beta_2) = g(\beta_2 - \beta_1)\int_{C_1}\frac{d\alpha_1}{2\pi}\int_{C_1}\frac{d\alpha_2}{2\pi} : e^{i\phi(\beta_1)+i\phi(\beta_2)-i\bar{\phi}(\alpha_1)-i\bar{\phi}(\alpha_2)} :$$
$$\times f(\alpha_1 - \beta_1)f(\alpha_2 - \beta_2)\bar{g}(\alpha_2 - \alpha_1)w(\alpha_2 - \beta_1)w(\beta_2 - \alpha_1). \qquad (3.11)$$

The factors $\bar{g}(\alpha_2 - \alpha_1), w(\alpha_2 - \beta_1), w(\beta_2 - \alpha_1)$ arise with the restriction

$$\mathrm{Im}(\alpha_2 - \alpha_1) < -\pi,\ \mathrm{Im}(\alpha_2 - \beta_1) < -\frac{\pi}{2},\ \mathrm{Im}(\beta_2 - \alpha_1) < -\frac{\pi}{2}$$

respectively. However, the location of the poles of these functions is consistent with the modification of the contours for $\alpha_1$ and $\alpha_2$ to $C_1$. Therefore, the equality (3.10) follows from the identity

$$S(f(\alpha_1 - \beta_1)f(\alpha_2 - \beta_2)\bar{g}(\alpha_2 - \alpha_1)w(\alpha_2 - \beta_1)w(\beta_2 - \alpha_1))$$
$$= S(f(\alpha_1 - \beta_1)f(\alpha_2 - \beta_2)\bar{g}(\alpha_1 - \alpha_2)w(\alpha_1 - \beta_2)w(\beta_1 - \alpha_2))$$

where $S$ means the symmetrization of $(\alpha_1, \alpha_2)$.

The proof of (2.35) is similar. In fact, it is even simpler because the contour $C_2$ is simpler than $C_1$. We omit the detail.

Let us consider (2.36). The case $a = b = 1$ follows immediately from (A.7). The case $a = -b = 1$ follows from (A.7), (A.8), (A.9). Note that the operator products (A.8), (A.9) produce no pole on the line $\text{Im}\,(\beta_2 - \beta_1) = 0$. Therefore, the contours in $Z'_-(\alpha)$ of the left and right hand sides are equal. The case $a = -b = -1$ is similar. Finally, consider the case $a = b = -1$. We have

$$Z_-(\beta_1)Z'_-(\beta_2) = -ie^{-\gamma}\int_{C_1}\frac{d\alpha_1}{2\pi}\int_{C_2}\frac{d\alpha_2}{2\pi} : e^{i\phi(\beta_1)-i\bar{\phi}(\alpha_1)+i\phi'(\beta_2)-i\bar{\phi}'(\alpha_2)} :$$
$$\times f(\alpha_1 - \beta_1)f'(\alpha_2 - \beta_2)h(\beta_2 - \beta_1)\frac{(\beta_2 - \alpha_1)(\alpha_2 - \beta_1)}{(\alpha_1 - \alpha_2)^2 + \frac{\pi^2}{4}}. \qquad (3.12)$$

The contour for $\alpha_1$ must go below $\beta_1 - \frac{\pi i}{2}$, while it must also avoid the pole at $\alpha_1 = \alpha_2 - \frac{\pi i}{2}$. Apparently there is a chance of pinching at $\text{Im}\beta_1 = 0$. However the contour for $\alpha_2$ can be deformed because $f'(\alpha_1 - \beta_2)$ has no pole if $\text{Im}(\alpha_2 - \beta_1) < \frac{\pi}{2}$. Therefore, the pinching does not occur. With a similar consideration for $Z'_-(\beta_2)Z_-(\beta_1)$ we obtain (2.36). □

**Proposition 3.2** *The operators* $Z_\pm(\beta)$ *and* $Z'_\pm(\beta)$ *satisfy (2.40).*

*Proof.* Note that in (A.7) the restriction to $\beta_2 - \beta_1 = -\frac{\pi i}{2}$ is regular. The case $a = b = +$ follows immediately from the identity

$$-\sinh\frac{\pi t\xi}{2}e^{i\beta t} + \sinh\frac{\pi t(\xi + 1)}{2}e^{i(\beta - \frac{\pi i}{2})t}$$
$$= \sinh\frac{\pi t(\xi + 1)}{2}e^{i(\beta - \pi i(\xi + \frac{1}{2}))t} - \sinh\frac{\pi t\xi}{2}e^{i(\beta - \pi i(\xi + 1))t}. \quad (3.13)$$

Next consider the case $a = -b = +$. We have an integration in $Z_-$. Call the integration variable $\alpha$. The equality of the integrands in the left and right hand sides follows from (3.13) and the identity

$$(\alpha - \beta)f(\alpha - \beta + \frac{\pi i}{2}) = f(\alpha - \beta + \pi i(\xi + \frac{1}{2}))(\beta - \alpha - \pi i(\xi + 1)).$$

The equality of the contours also follows because the irrelevant poles are cancelled by the factors $\alpha - \beta$ and $\beta - \alpha - \pi i(\xi + 1)$. The case $a = -b = -$ is similar.

In the case $a = b = -$, the only additional care we should take is to choose the integration contour for the variable $\alpha_1$ in the factors $Z'_-(\beta)$, $Z'_-(\beta^* - \frac{\pi i}{2})$, and $C_2$ for $\alpha_2$ in $Z_-(\beta - \frac{\pi i}{2})$, $Z_-(\beta^*)$ in such a way that $C_1$ is above $\alpha_2 + \frac{\pi i}{2}$ in the left hand side, and below $\alpha_2 - \frac{\pi i}{2}$ in the right hand side. This is possible because the pole of $f(\alpha_2 - \beta + \frac{\pi i}{2})$ at $\alpha_2 = \beta$ and the pole of $f(\alpha_2 - \beta + \pi i(\xi + \frac{1}{2}))$ at $\alpha_2 = \beta - \pi i(\xi + 1)$ are cancelled.   □

**Proposition 3.3** *The operators $Z_\pm(\beta)$ satisfy (2.38). Likewise $Z'_\pm(\beta)$ satisfy (2.39). The constants $C, C'$ are given by*

$$C = c_1 \sin\frac{\pi}{\xi}\left(\pi\xi\Gamma(-\frac{1}{\xi})\right)^2 g(-\pi i), \qquad c_1 = \frac{1}{\pi i}e^{-(\gamma+\log(\pi\xi))(\xi+1)/\xi},$$

$$C' = c'_1 \sin\frac{\pi}{\xi+1}\left(\pi(\xi+1)\Gamma(\frac{1}{\xi+1})\right)^2 \lim_{\beta\to 0}\frac{g'(\beta+\pi i)}{\beta},$$

$$c'_1 = \frac{1}{\pi i}e^{-(\gamma+\log(\pi(\xi+1)))\xi/(\xi+1)}.$$

*Proof.* We shall show the case of $Z_\pm(\beta)$.

Since $g(\beta_1 - \beta_2)$ is regular at $\beta_1 = \beta_2 + \pi i$, $Z_+(\beta_1)Z_+(\beta_2)$ is also regular there. Consider $Z_+(\beta_1)Z_-(\beta_2)$ given in (3.7). As $\beta_1$ tends to $\beta_2 + \pi i$, the contour is pinched. Taking the residue at $\alpha = \beta_1 - \pi i/2$ we find that modulo regular terms the right hand side is written as

$$c_1 g(\beta_2 - \beta_1) f(\beta_1 - \beta_2 - \frac{\pi i}{2})\frac{\pi\xi}{i}\sin\frac{\pi}{\xi}\Gamma\left(-\frac{1}{\xi}\right) : e^{-i\phi(\beta_1-\pi i)+i\phi(\beta_2)} :.$$

Here we have used

$$\bar\phi(\beta) = \phi(\beta + \frac{\pi i}{2}) + \phi(\beta - \frac{\pi i}{2}).$$

The behavior at $\beta_1 = \beta_2 + \pi i$ is easily calculated from this. The case of $Z_-(\beta_1)Z_+(\beta_2)$ is similar.

Finally consider $Z_-(\beta_1)Z_-(\beta_2)$. In the right hand side of (3.11), we rewrite the integral into three parts as

$$I_1 + I_2 + I_3 = \int_{C_{\beta_1-\pi i/2}}\frac{d\alpha_1}{2\pi}\int_{C_1}\frac{d\alpha_2}{2\pi} + \int_{\tilde C_1}\frac{d\alpha_1}{2\pi}\int_{C_{\beta_1-\pi i/2}}\frac{d\alpha_2}{2\pi} + \int_{\tilde C_1}\frac{d\alpha_1}{2\pi}\int_{\tilde C_1}\frac{d\alpha_2}{2\pi},$$

where $C_{\beta_1-\pi i/2}$ is a small circle around $\beta_1 - \pi i/2$, and $\tilde C_1$ is the same as $C_1$ except that $\beta_1 - \pi i/2$ is below $\tilde C_1$. Clearly $I_3$ is regular. Taking the residues

we find

$$\begin{aligned}
I_1 &= Af(\beta_1 - \beta_2 - \pi i/2)\sinh\frac{\beta_2 - \beta_1}{\xi} \\
&\quad \times \int_{C_1} \frac{d\alpha_2}{2\pi} \frac{f(\alpha_2 - \beta_2)}{w(\alpha_2 - \beta_1 + \pi)} :e^{-i\phi(\beta_1 - \pi i) + i\phi(\beta_2) - i\bar\phi(\alpha_2)}:, \\
I_2 &= Af(\beta_1 - \beta_2 - \pi i/2)\sinh\frac{-\pi i}{\xi} \\
&\quad \times \int_{\tilde C_1} \frac{d\alpha_1}{2\pi} \frac{f(\alpha_1 - \beta_1)w(\beta_2 - \alpha_1)}{w(\alpha_1 - \beta_1 - \pi)w(\beta_1 - \alpha_1)} :e^{-i\phi(\beta_1 - \pi i) + i\phi(\beta_2) - i\bar\phi(\alpha_1)}:,
\end{aligned}$$

where $A$ is a constant. In the above, the integral part for $I_1$ is now regular because of the zero of the integrand. The pole at $\beta_1 = \beta_2 + \pi i$ comes from $f(\beta_1 - \beta_2 - \pi i/2)$. Taking the residue we find that the residues for $I_1$ and $I_2$ cancel each other. This completes the proof. $\square$

A similar calculation shows the following.

**Proposition 3.4**

$$\mathrm{Res}_{\beta_1 = \beta_2 + \pi i - \pi i \xi} Z_a(\beta_1) Z_b(\beta_2) = const.\delta_{a+b,0} T(\beta),$$
$$T(\beta) =: e^{\lambda(\beta + \pi i/2)}: + :e^{-\lambda(\beta - \pi i/2)}:,$$

*where*

$$\lambda(\beta) = i\phi(\beta - \frac{\pi i}{2}) - i\phi(\beta - \frac{\pi i}{2} + \pi i\xi) = \int \frac{2\sinh(\pi\xi t/2)}{\sinh \pi t} e^{it\beta + \pi(\xi+1)t/2} a(t) dt.$$

The operator $T(\beta)$ is nothing but the limit of the current of the deformed Virasoro algebra introduced by Shiraishi et al. [12]. This agrees with Lukyanov's observation [13] that the deformed Virasoro algebra (in the limit) and the creation operators of breathers in sine-Gordon theory satisfy the same commutation relations.

*Remark*  The bosonization given above is basically obtained from Lukyanov's [6] by setting $t = m\varepsilon$, $a(t) = a_m/\varepsilon$ and letting $\varepsilon \to 0$. This procedure gives rise to the following expressions for the vertex operators

$$Z_+(\beta) = e^{-2\beta/\xi} :e^{i\varphi(\beta)}:, \tag{3.14}$$

$$Z_-(\beta) = e^{2\beta/\xi} \int \frac{d\alpha}{2\pi} e^{(\alpha-\beta)/\xi} f(\alpha - \beta) :e^{i\varphi(\beta) - i\bar\varphi(\alpha)}:, \tag{3.15}$$

and similar formulars for $Z'_\pm(\beta)$. Here

$$i\varphi(\beta) = \sqrt{\frac{\xi+1}{2\xi}} iQ + i\phi(\beta), \qquad i\bar\varphi(\beta) = i\varphi(\beta - \frac{\pi i}{2}) + i\varphi(\beta + \frac{\pi i}{2}).$$

They differ from the formulas (3.3),(3.4) by the presence of a piece $Q$ of the 'zero mode' operators, and an exponential factor. They satisfy the same commutation relations (2.34)-(2.36). However we have not been able to recover the formulas for the correlation functions (1.3) using (3.14),(3.15) and their analogs. For this reason we modified the operators as given in (3.3)-(3.6).

## 4. Massless correlation functions

In this section, we derive a solution of the difference equations (1.4)-(1.6) algebraically, and obtain an integral formula for it.

Let us define the boost operator $H$ as

$$H = \int_0^\infty dt \frac{t^2 \sinh\frac{\pi t(\xi+1)}{2}}{\sinh\frac{\pi t}{2}\sinh\pi t \sinh\frac{\pi t\xi}{2}} a'(-t)a'(t) \tag{4.1}$$

which enjoys the property

$$e^{\lambda H} a'(t) e^{-\lambda H} = e^{-\lambda t} a'(t). \tag{4.2}$$

Hence we have

$$e^{\lambda H} X(\beta) e^{-\lambda H} = X(\beta + i\lambda), \tag{4.3}$$

for $X = V, \bar{V}, V', \bar{V}'$. Now let us consider the following trace function.

$$G(\beta_1, \cdots, \beta_{2n})_{\varepsilon_1, \cdots, \varepsilon_{2n}} = \frac{\text{tr}_{\mathcal{H}}(e^{-\lambda H} Z'_{\varepsilon_1}(\beta_1) \cdots Z'_{\varepsilon_{2n}}(\beta_{2n}))}{\text{tr}_{\mathcal{H}}(e^{-\lambda H})}. \tag{4.4}$$

By using the relations (2.35) and (4.3), one can show that the function $G(\beta_1, \cdots, \beta_{2n})_{\varepsilon_1, \cdots, \varepsilon_{2n}}$ satisfies the desired difference equations (1.4)-(1.6).

The bosonization formulas (3.5) and (3.6) for the vertex operators make the evaluation of the trace in (4.4) possible. Here we consider the basic function $G(\beta_1, \cdots, \beta_n, \beta_{n+1} \cdots, \beta_{2n})_{+\cdots+-\cdots-}$. The general components are obtainable from this by the relation (1.4).

From (3.5) and (3.6), we have

$$G(\beta_1, \cdots, \beta_n, \beta_{n+1} \cdots, \beta_{2n})_{+\cdots+-\cdots-}$$
$$= \frac{1}{\text{tr}_{\mathcal{H}}(e^{-\lambda H})} \int \prod_{a=n+1}^{2n} \left(\frac{d\alpha_a}{2\pi} f'(\beta_a - \alpha_a)\right)$$
$$\times \text{tr}_{\mathcal{H}}(e^{-\lambda H} V'(\beta_1) \cdots V'(\beta_n) : e^{i\phi(\beta_{n+1})-i\bar{\phi}(\alpha_{n+1})} : \cdots : e^{i\phi(\beta_{2n})-i\bar{\phi}(\alpha_{2n})} :).$$
$$\tag{4.5}$$

Carrying out a normal ordering procedure, we obtain

$$V'(\beta_1)\cdots V'(\beta_n) : e^{i\phi(\beta_{n+1})-i\bar\phi(\alpha_{n+1})} : \cdots : e^{i\phi(\beta_{2n})-i\bar\phi(\alpha_{2n})} :$$
$$= \prod_{1\le j<k\le 2n} g'(\beta_k-\beta_j) \prod_{1\le j\le 2n} \prod_{n+1\le a\le 2n} w'(\alpha_a-\beta_j)$$
$$\times \prod_{n+1\le a<b\le 2n} w'(\alpha_b-\beta_a)w'(\beta_b-\alpha_a)\bar g'(\alpha_b-\alpha_a)$$
$$\times e^{\int_0^\infty dt\mathcal{F}(-t)a'(-t)} e^{\int_0^\infty dt\mathcal{F}(t)a'(t)}, \qquad (4.6)$$

where

$$\mathcal{F}(t) = -\sum_{j=1}^{2n} \frac{e^{i\beta_j t}}{\sinh\pi t} + \sum_{a=n+1}^{2n} \frac{e^{i\alpha_a t}}{\sinh\frac{\pi t}{2}}. \qquad (4.7)$$

Then the trace is evaluated by using the following formula.

$$\operatorname{tr}_\mathcal{H}(e^{-\lambda H} e^{\int_0^\infty dt g(t)a'(-t)} e^{\int_0^\infty dt h(t)a'(t)})/\operatorname{tr}_\mathcal{H}(e^{-\lambda H})$$
$$= \exp\Bigl\{\int_0^\infty dt \frac{\sinh\frac{\pi t}{2}\sinh\pi t\sinh\frac{\pi t\xi}{2}}{t\sinh\frac{\pi t(\xi+1)}{2}(e^{\lambda t}-1)} g(t)h(t)\Bigr\} \qquad (4.8)$$

for some functions $g(t)$ and $h(t)$.

Set $\nu = 1/(\xi+1)$. Using this formula in (4.5), we have the following result [1].

$$G(\beta_1,\cdots,\beta_n,\beta_{n+1}\cdots,\beta_{2n})_{+\cdots+-\cdots-}$$
$$= \prod_{1\le j<k\le 2n} \rho(\beta_j-\beta_k) \prod_{a=n+1}^{2n} \int_{\mathcal{C}_a} \frac{d\alpha_a}{2\pi} \prod_{j=1}^{2n}\prod_{a=n+1}^{2n} \varphi(\beta_j-\alpha_a)$$
$$\times \prod_{n+1\le a<b\le 2n} \frac{\psi(\alpha_a-\alpha_b)}{\sinh\nu(\alpha_a-\alpha_b-\pi i)}$$
$$\times \prod_{\substack{j<a\\1\le j\le 2n\\n+1\le a\le 2n}} \sinh\nu(\alpha_a-\beta_j+\frac{\pi i}{2}) \prod_{n+1\le a<j\le 2n} \sinh\nu(\beta_j-\alpha_a+\frac{\pi i}{2})$$
$$\qquad (4.9)$$

Here we omitted an irrelevant constant factor. The functions $\rho(\beta),\varphi(\beta)$ and $\psi(\alpha)$ are defined by

$$\rho(\beta) = \sinh\frac{\pi\beta}{\lambda} \frac{S_3(\pi-i\beta|\lambda,2\pi,\frac{\pi}{\nu})S_3(\pi+\lambda+i\beta|\lambda,2\pi,\frac{\pi}{\nu})}{S_3(-i\beta|\lambda,2\pi,\frac{\pi}{\nu})S_3(\lambda+i\beta|\lambda,2\pi,\frac{\pi}{\nu})}, \quad (4.10)$$

$$\varphi(\beta) = \frac{2}{S_2(\frac{\pi}{2}+i\beta|\lambda,\frac{\pi}{\nu})S_2(\frac{\pi}{2}-i\beta|\lambda,\frac{\pi}{\nu})}, \qquad (4.11)$$

$$\psi(\alpha) = \sinh\frac{\pi\beta}{\lambda} S_2(\pi+i\beta|\lambda,\frac{\pi}{\nu})S_2(\pi-i\beta|\lambda,\frac{\pi}{\nu}). \qquad (4.12)$$

For the definition of the multiple sine function $S_r(x|\omega_1,\cdots,\omega_r)$, see Appendix A of [1].

In the derivation of (4.9), we made the following identification.

$$g'(\beta_k - \beta_j)\exp\left\{-\int_0^\infty dt \frac{\sinh\frac{\pi t}{2}\sinh\frac{\pi t\xi}{2}}{t\sinh\pi t\sinh\frac{\pi t(\xi+1)}{2}} \frac{e^{i(\beta_j-\beta_k)t}+e^{-i(\beta_j-\beta_k)t}}{e^{\lambda t}-1}\right\}$$
$$= \text{const.}\rho(\beta_j - \beta_k), \tag{4.13}$$

$$\bar{g}'(\alpha_b - \alpha_a)\exp\left\{-\int_0^\infty dt \frac{\sinh\pi t\sinh\frac{\pi t\xi}{2}}{t\sinh\frac{\pi t}{2}\sinh\frac{\pi t(\xi+1)}{2}} \frac{e^{i(\alpha_a-\alpha_b)t}+e^{-i(\alpha_a-\alpha_b)t}}{e^{\lambda t}-1}\right\}$$
$$= \text{const.}\frac{\psi(\alpha_a - \alpha_b)}{\sinh\nu(\alpha_a - \alpha_b - \pi i)}, \tag{4.14}$$

$$w'(\alpha - \beta)\exp\left\{\int_0^\infty dt \frac{\sinh\frac{\pi t\xi}{2}}{t\sinh\frac{\pi t(\xi+1)}{2}} \frac{e^{i(\beta-\alpha)t}+e^{-i(\beta-\alpha)t}}{e^{\lambda t}-1}\right\}$$
$$= \text{const.}\varphi(\beta - \alpha)\sinh\nu(\alpha - \beta + \frac{\pi i}{2}). \tag{4.15}$$

The resultant expression (4.9) coincides with the one obtained in [1] previously.

**Acknowledgements**

Two of us (M.J. and T.M.) are indebted to the organizers of the workshop, in particular to Moshe Flato and Daniel Sternheimer, for kind invitation and hospitality. We thank S. Khoroshkin, D. Lebedev, S. Lukyanov, S. Pakuliak and Y. Pugai for stimulating discussions.

**A. Appendix**

Here we list the formulas of the form

$$X(\beta_1)Y(\beta_2) = C_{X,Y}(\beta_2 - \beta_1) : X(\beta_1)Y(\beta_2) :$$

where $X, Y = V, \bar{V}, V', \bar{V}'$ and $C_{X,Y}(\beta)$ is a meromorphic function on $\mathbf{C}$. The equality $C_{X,Y} = C_{Y,X}$ is valid in all cases. Therefore, we omit redundant cases.

These formulas follow from the definition of $V(\alpha)$, $\bar{V}(\alpha)$, $V'(\alpha)$, $\bar{V}'(\alpha)$ and the commutation relation (3.1). We first obtain $\log C_{X,Y}$ as an integral of the form:

$$\int_0^\infty f(t)dt = \int_C f(t)\frac{\log(-t)dt}{2\pi i}$$

where the regularization prescription (3.2) is implied. The resulting integrals can be written by using the gamma function $\Gamma(x)$ or the double gamma function $\Gamma_2(x|\omega_1,\omega_2)$. For a summary of definitions and properties of the latter, see Appendix A in [1]. In each formula, we have shown the region of the absolute convergence of the integral by $\mathrm{Im}(\beta_2 - \beta_1) < 0$, etc. This information is useful when we fix the integration contours in the formulas (3.4) and (3.6).

$$V(\beta_1)V(\beta_2) = g(\beta_2 - \beta_1) : e^{i\phi(\beta_1)+i\phi(\beta_2)} : \quad (\mathrm{Im}(\beta_2 - \beta_1) < 0) \tag{A.1}$$

$$g(\beta) = \frac{\Gamma_2(i\beta + \pi|2\pi,\pi\xi)\Gamma_2(i\beta + \pi(\xi+1)|2\pi,\pi\xi)}{\Gamma_2(i\beta|2\pi,\pi\xi)\Gamma_2(i\beta + \pi(\xi+2)|2\pi,\pi\xi)} e^{\frac{\gamma(\xi+1)}{2\xi}}$$

$$V(\beta_1)\bar{V}(\beta_2) = w(\beta_2 - \beta_1) : e^{i\phi(\beta_1)-i\bar{\phi}(\beta_2)} : \quad (\mathrm{Im}(\beta_2 - \beta_1) < -\frac{\pi}{2}) \tag{A.2}$$

$$w(\beta) = \frac{\Gamma(\frac{i\beta}{\pi\xi} - \frac{1}{2\xi})}{\Gamma(\frac{i\beta}{\pi\xi} + 1 + \frac{1}{2\xi})} e^{-(\gamma+\log(\pi\xi))\frac{\xi+1}{\xi}} = \frac{1}{g(\beta + \frac{\pi i}{2})g(\beta - \frac{\pi i}{2})}$$

$$\bar{V}(\beta_1)\bar{V}(\beta_2) = \bar{g}(\beta_2 - \beta_1) : e^{-i\bar{\phi}(\beta_1)-i\bar{\phi}(\beta_2)} : \quad (\mathrm{Im}(\beta_2 - \beta_1) < -\pi) \tag{A.3}$$

$$\bar{g}(\beta) = \frac{i\beta}{\pi\xi}\frac{\Gamma(\frac{i\beta}{\pi\xi} + 1 + \frac{1}{\xi})}{\Gamma(\frac{i\beta}{\pi\xi} - \frac{1}{\xi})} e^{2(\gamma+\log(\pi\xi))\frac{\xi+1}{\xi}} = \frac{1}{w(\beta + \frac{\pi i}{2})w(\beta - \frac{\pi i}{2})}$$

$$V'(\beta_1)V'(\beta_2) = g'(\beta_2 - \beta_1) : e^{i\phi'(\beta_1)+i\phi'(\beta_2)} : \quad (\mathrm{Im}(\beta_2 - \beta_1) < \pi) \tag{A.4}$$

$$g'(\beta) = \frac{\Gamma_2(i\beta + 2\pi|2\pi,\pi(\xi+1))\Gamma_2(i\beta + \pi(\xi+1)|2\pi,\pi(\xi+1))}{\Gamma_2(i\beta + \pi|2\pi,\pi(\xi+1))\Gamma_2(i\beta + \pi(\xi+2)|2\pi,\pi(\xi+1))} e^{\frac{\gamma\xi}{2(\xi+1)}}$$

$$V'(\beta_1)\bar{V}'(\beta_2) = w'(\beta_2 - \beta_1) : e^{i\phi'(\beta_1)-i\bar{\phi}'(\beta_2)} : \quad (\mathrm{Im}(\beta_2 - \beta_1) < \frac{\pi}{2}) \tag{A.5}$$

$$w'(\beta) = \frac{\Gamma(\frac{i\beta}{\pi(\xi+1)} + \frac{1}{2(\xi+1)})}{\Gamma(\frac{i\beta}{\pi(\xi+1)} + 1 - \frac{1}{2(\xi+1)})} e^{-(\gamma+\log\pi(\xi+1))\frac{\xi}{\xi+1}} = \frac{1}{g'(\beta + \frac{\pi i}{2})g'(\beta - \frac{\pi i}{2})}$$

$$\bar{V}'(\beta_1)\bar{V}'(\beta_2) = \bar{g}'(\beta_2 - \beta_1) : e^{-i\bar{\phi}'(\beta_1)-i\bar{\phi}'(\beta_2)} : \quad (\mathrm{Im}(\beta_2 - \beta_1) < 0) \tag{A.6}$$

$$\bar{g}'(\beta) = \frac{i\beta}{\pi(\xi+1)}\frac{\Gamma(\frac{i\beta}{\pi(\xi+1)} + 1 - \frac{1}{\xi+1})}{\Gamma(\frac{i\beta}{\pi(\xi+1)} + \frac{1}{\xi+1})} e^{2(\gamma+\log\pi(\xi+1))\frac{\xi}{\xi+1}} = \frac{1}{w'(\beta + \frac{\pi i}{2})w'(\beta - \frac{\pi i}{2})}$$

$$V(\beta_1)V'(\beta_2) = h(\beta_2 - \beta_1) : e^{i\phi(\beta_1)+i\phi'(\beta_2)} : \quad (\mathrm{Im}(\beta_2 - \beta_1) < \frac{\pi}{2}) \tag{A.7}$$

$$h(\beta) = \frac{\Gamma(\frac{i\beta}{2\pi} + \frac{1}{4})}{\Gamma(\frac{i\beta}{2\pi} + \frac{3}{4})} e^{-\frac{1}{2}(\gamma+\log(2\pi))}$$

$$V(\beta_1)\bar{V}'(\beta_2) = i(\beta_2 - \beta_1)e^\gamma : e^{i\phi(\beta_1)-i\bar{\phi}'(\beta_2)} : \quad (\mathrm{Im}(\beta_2 - \beta_1) < 0) \tag{A.8}$$

$$\bar{V}(\beta_1)V'(\beta_2) = i(\beta_2 - \beta_1)e^\gamma : e^{-i\bar{\phi}(\beta_1)+i\phi'(\beta_2)} : \quad (\mathrm{Im}(\beta_2 - \beta_1) < 0) \tag{A.9}$$

$$\bar{V}(\beta_1)\bar{V}'(\beta_2) = -\frac{e^{-2\gamma}}{(\beta_2 - \beta_1)^2 + \frac{\pi^2}{4}} : e^{-i\bar{\phi}(\beta_1)-i\bar{\phi}'(\beta_2)} : \quad (\mathrm{Im}(\beta_2 - \beta_1) < -\frac{\pi}{2}) \tag{A.10}$$

We set
$$S_0(\beta) = \frac{g(-\beta)}{g(\beta)}, \qquad R_0(\beta) = \frac{g'(-\beta)}{g'(\beta)}. \tag{A.11}$$

The following relations are sometimes useful.

$$\frac{w(\beta)}{w(-\beta)} = -\frac{\sinh(\beta - \frac{\pi i}{2})/\xi}{\sinh(\beta + \frac{\pi i}{2})/\xi}, \tag{A.12}$$

$$\frac{\bar{g}(\beta)}{\bar{g}(-\beta)} = \frac{\sinh(\beta + \pi i)/\xi}{\sinh(\beta - \pi i)/\xi}, \tag{A.13}$$

$$\frac{w'(\beta)}{w'(-\beta)} = -\frac{\sinh\frac{\beta + \frac{\pi i}{2}}{\xi+1}}{\sinh\frac{\beta - \frac{\pi i}{2}}{\xi+1}}, \tag{A.14}$$

$$\frac{\bar{g}'(\beta)}{\bar{g}'(-\beta)} = \frac{\sinh\frac{\beta - \pi i}{\xi+1}}{\sinh\frac{\beta + \pi i}{\xi+1}}, \tag{A.15}$$

$$\frac{h(\beta)}{h(-\beta)} = -\frac{\sinh(\frac{\beta}{2} - \frac{\pi i}{4})}{\sinh(\frac{\beta}{2} + \frac{\pi i}{4})}. \tag{A.16}$$